# Simulating Keystroke and Computing the Theoretical Probability of Infinite Monkey Theorem with Markov Process

Juncheng Yi [1*], Hongyi Jiang[2], and Kaiwen Zhou[3]

[1]College of Arts & Sciences, University of Washington, Seattle, United States

[2]College of Arts & Sciences, University of Washington, Seattle, United States

[3]College of Arts & Sciences, University of Washington, Seattle, United States

*Corresponding author: ccyi3@iastate.edu

**Abstract.** /The Infinite Monkey Theorem states that if one monkey randomly hits the keys in front of a typewriter keyboard during an infinite amount of time, any works written by William Shakespeare will almost surely be typed out at the end of the total text. Due to the seemingly low chance of typing the exact literature works, our group are motivated to find out the expected time the *Hamlet*, our target text, being typed out by simulated random typing on a standard keyboard. For finding the answer, 30 users randomly typed characters into a file. Then, the frequency of each characters occurred following the previous character is calculated. This conditional probability is used to build the Markov matrix by considering all 128×128 cases. Finally, the expected time we estimated is $(10^{134}) \times (\frac{78}{760})$ min, which is surprisingly lower than the theoretical computation, and not achievable at all even in the cosmic time.

## 1. Introduction

### 1.1 Problem Description

The infinite monkey theorem states that a monkey hitting keys at random on a typewriter keyboard for an infinite amount of time will type any given text with probability 1, such as the Hamlet Poem by William Shakespeare[1]. Since the "monkey" metaphor first talked in 1913, it gradually became more attractive to ordinary people that they would get a sense of how mathematical probability works and why some unlikely events will theoretically happen with probability 1 in the long run.

In 2002, experimenters from to University of Plymouth tested the actual situation of this theorem: they tried inviting a group of actual monkeys to stay with a typewriter keyboard, but finally the monkeys damaged the keyboard, and the experimenters only received some papers with nearly all "S"[2]. As a result, they conclude that the infinite monkey theorem is far more complicated with more factors in actual situations and the probability to type Hamlet perfect in the real case is even lower than it in the theoretical case[2].

Inspired by this 2002 experiment, we would like to calculate how small the probability is if someone like monkeys just randomly hits the typewriter keyboard in ideal condition: continuing hitting some keys on the keyboard without breaking it in an infinite time, until the same *Hamlet* poem is printed out. We would like to try one new algorithm to calculate this probability, discuss its limitations and compare its results to other algorithms' conclusions. By the way, we also want to see some typing patterns on the keyboard from humans and how it is different from the common English text by visualization.

Thus, the objectives of this research are:

(a). Calculate the probability and the average time for a person who randomly clicks the keyboard to correctly type Shakespeare's poem *Hamlet*. (Markov matrix to match the graph pattern)

(b). Visualize people's typing patterns on an English keyboard in 2022. (GeoPandas)

This research is structured as follow:

1. Data Sampling
2. Discussion on analysis approach
3. Build Markov Matrix
4. PCA application
5. Build Markov Matrix with rational number
6. Compute text probability

### 1.2 Literature Review

Due to the impracticality of testing the theorem with monkeys in reality,  researchers mainly tested Infinite Monkey Theorem through computer simulations. One of the most popular methods to calculate the probability of a string is by assuming that each character in the string is independent from another and has the equal chance being typed by a monkey[3]. Another interesting simulation method is to foster a set of monkeys who have an equal chance of clicking each character at first[4]. Then, the frequency of each letter being hitten changes in every iteration[4]. By setting the monkey with a higher chance to find words as the parent monkey, its letter frequency would be "passed down to its child monkey"[4].  Besides, other than considering the theorem as strings, Banerji, Mansour, and Severini followed the notion of graph likelihoods. In this case, they reframed the question by asking the probability of a monkey constructing any given graphs with specified vertices[5].

Inspired by former research, our group simulates the first character according to the uniform distribution, because no strong dependency relationship has been shown yet. Inspired by the graph likelihood approach, our group takes consideration of the dependency of characters on the keyboard. Thus, all characters except for the first one are simulated according to the dependency shown in our collected data.

## 2. Theoretical Background

### 2.1 Discrete Cumulative Distribution Function

The Cumulative Distribution Function (CDF) of random variable X evaluated at $x$ gives the probability of X taking on a value less than or equal to $x$[6]. The function is written as follows.

$$F_x(x) = P(X \leq x)$$

The main properties of the CDF are listed here[6]:

1. The function is right-continuous monotone increasing.
2. $\lim_{x \to -\infty} F(x) = 0$
3. $\lim_{x \to \infty} F(x) = 1$

In this paper, we use Discrete CDF as the main method for the character's generating simulation.

## 2.2 Markov (Stochastic) Chain

Markov Chain describes a sequence of possible events in which the probability of each event depends only on the previous event[7]. For example, someone who just hits "k" on the keyboard may be more likely to hit keys such as "j", "i" and "l" that are close to "k" when randomly hitting the keys. Thus, rather than assuming the probability of hitting each key is equal and independent, our group decides to use the Markov Chain as a model.

Bolzano-Weierstrass convergence theorem guarantees convergence when we apply the same mapping repeatedly regardless of the input. After reaching the steady state, it forgets about the input if only the current state is stored, resulting in a large-converged-vector histogram.

We can apply Bolzano-Weierstrass convergence theorem here to say, it's doomed to converge, if you apply this mapping repeatedly over a start result, regardless of input. After reaching steady state, it forgets about where it came from if only the current state is stored, and will give a very large converged-vector histogram. (We can go deep into this) So, we should limit the sample size.

**Proof**: BW and Banach contraction space!(Caution, weak solution because of below assumption)
We first understand {M^i x}_i=1 ^ +inf as a sequence. This is a seq of operations by repeatedly applying M on any initial x in R128.
We also borrow the fact that for all Markov matrices, there is an eigenvalue = 1 and for all other eigenvalues, this value's absolute value is smaller than 1.
Assumption: this markov matrix has full rank, all different eigenvalues and thus all linear independent eigenvectors. We have already normalized the eigen basis, so all eigen vector has value of 1.

This sequence belongs to R128 and R 128 is bounded, so this sequence is bounded.
This sequence is bounded implies there is an accumulation point s.t. Mpt = pt. This is exactly the principal value, i.e. the scaled eigenvector corresponding to unique eigenvalue 1.
Accumulation point is good. Now go for the contraction.
We want for all x in R128, |T(x) - T(pt)|/|x - pt| < 1. This is easy because all x in R 128 can be written in eigen-decomposition form EV1*C1 + EV2*C2 + … + EV128*C128.
So, T(x) - T(pt) = A(x - pt) = (C1 - 1)*EV1*1 + C2*EV2*LAMBDA2 + … + C128*EV128*LAMBDA128 where all lambdas != 1 are smaller than 1 in magnitude.
Then x - pt = EV1*(C1-1) + EV2*C2 + … + EV128*C128
Notice that the vectors are just L.I., not orthogonal, so we will need a gram-schmidt process to fully explain the following phenomena. We then got x-pt gram-schmidt-ed, align with EV1. We have new basis

EV1, ONB2, ONB3, …, ONBn. They will be equivalent to EV1, (EV2 - R21*EV1)/norm(EV2 - R21*EV1), (EV3 - R32*(EV2 - R21*EV1)/norm(EV2 - R21*EV1)) - R31*EV1…
We then have the following lemma: For all n, the coefficient NCn before ONn has an upper bound of Cn. This is caused by the operation essence of gram-schmidt, which is just removing previous components from a vector. If this vector is already orthogonal to all vectors in front, then this NCn = Cn. Otherwise, NCn = Cn - Sigma(j, proj(EVn, prevEVj)). This is an application of Cauchy Schwarz.
So the maximum case ratio would be (|EV1|**2 * (C1 - 1)**2 + |EV2|**2*C2**2*LAMBDA2**2 + … + |EV128|**2*C128**2*LAMBDA128**2)/ (|EV1|**2 * (C1 - 1)**2 + |EV2|**2*C2**2 + … + |EV128|**2*C128**2), which is absolutely smaller than 1 since all lambdas other than lambda1 are smaller than 1 in magnitude. This squeezes all general cases.
Q.E.D.

### 2.3 Singular Value Decomposition and Principal component analysis

We apply principal component analysis (PCA) on the Markov matrix here for the convergence vector. PCA is often used for analyzing multidimensional data[8]. It's also a useful technique to reduce dimensionality of the dataset[8].
We make following ruleset for eigenvalue and vector handling:

1. If there's an eigenvalue way larger than 1, the model is problematic.
2. If there's an eigenvalue almost 1, keep it and its corresponding eigenvector. This is a convergence vector for sure.
3. If there is an eigenvalue smaller than 1, discard it and its corresponding eigenvector if the eigenvalue is smaller than 1.

In implementation, we use Singular Value Decomposition (SVD) for the following properties:

1. A symmetric matrix always gives orthogonal eigenvectors.
2. sum, determinant, offset are all facilitated by symmetric properties.

## 3. Data Collection and Analysis

### 3.1 Overview

In order to build a matrix that models the transition between each "small step" — adding one additional character based on the current character sequence, first we need something to fill in each entry. Therefore, we will record people's randomly-typing results on a typewriter keyboard, and calculate the frequency and conditional probability based on the record — the text sample. These conditional probability and frequency will be filled into the entries of the transition matrix later.

### 3.2 Defining Terms

Because different keyboards of different device model designed by different companies is different in size, number and relative position of keys, and some other features, in order to make data collection process from keyboard consistent, we need to define a "Standard Typewriter Keyboard" that is used by all of the process in this research. There are two requirements for a Standard Typewriter Keyboard:
1. It should have all keys or combinations of keys that produce common characters in text.
2. It should be able to produce all characters in Shakespeare's *Hamlet*.

Therefore, we decided to use LG Rog Strix Flare as “Standard Typewriter Keyboard”.

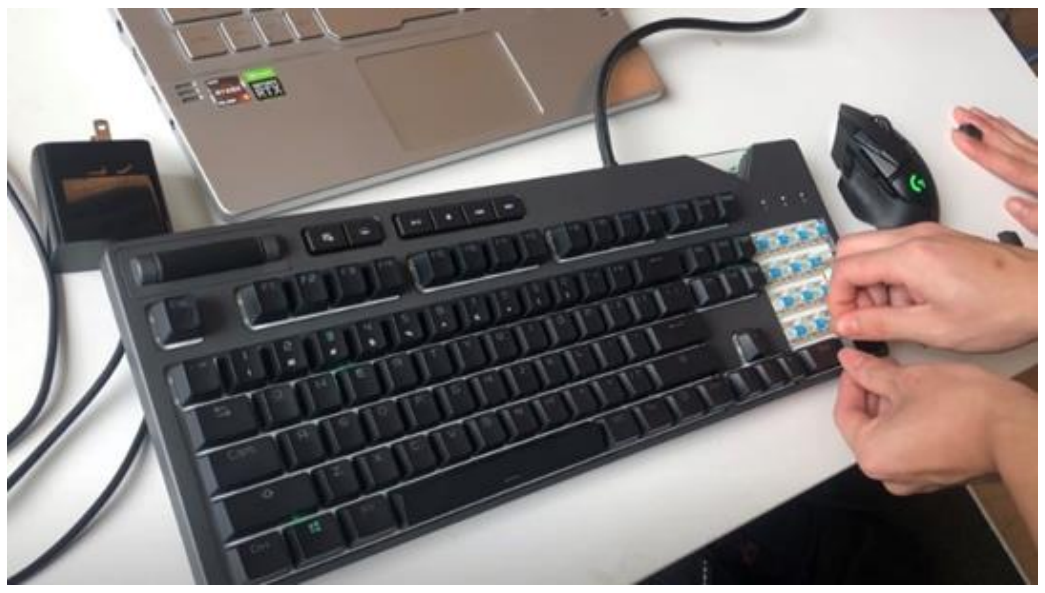

Removing unnecessary keycaps on LG Rog Strix Flare

In addition, we discovered that if someone hits some combination of function keys on the keyboard, such as “Esc”, “Fn”, “Alt”, … the file will terminate accidentally and thus the typed text will not be saved. We thus eliminate this possibility to make the condition more ideal. The corresponding problem is solved by ARMOURY CRATE, which allows us to remove the functionality of specified keys. The remaining keys are colored in white and the removed keys are colored in red.

Two versions of “Standard Typewriter Keyboard” are defined:
1. Testing Simple Version: with only the 26 English letters’ keys (for testing the algorithm’s operation results)

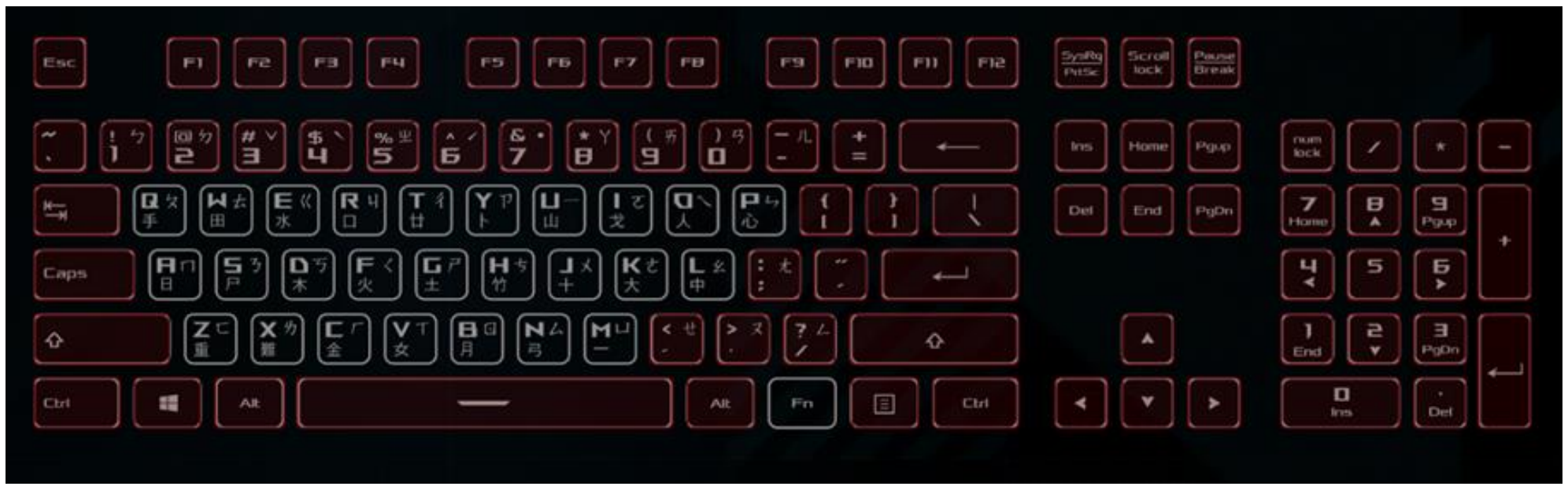

2. Realistic Version: with all common keys or combinations of keys that can result at least one character on the typewriter keyboard (for setting up the sample space for the research’s final findings)

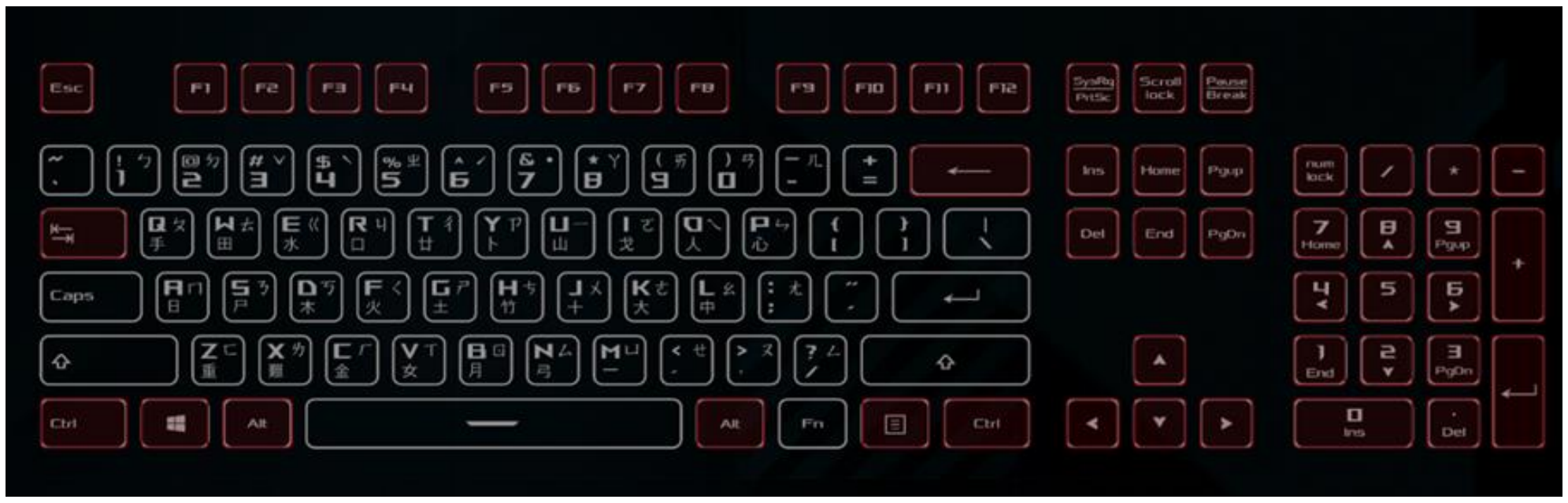

In order to calculate the probability of typing Shakespeare’s *Hamlet* correctly, defining the *Hamlet* is important because the poem is famous and has many different versions. We define the *“Hamlet”* in our research strictly as the *hamlet* in GitHub. We download and name it as “hamlet.txt” in the sample folders [9]:

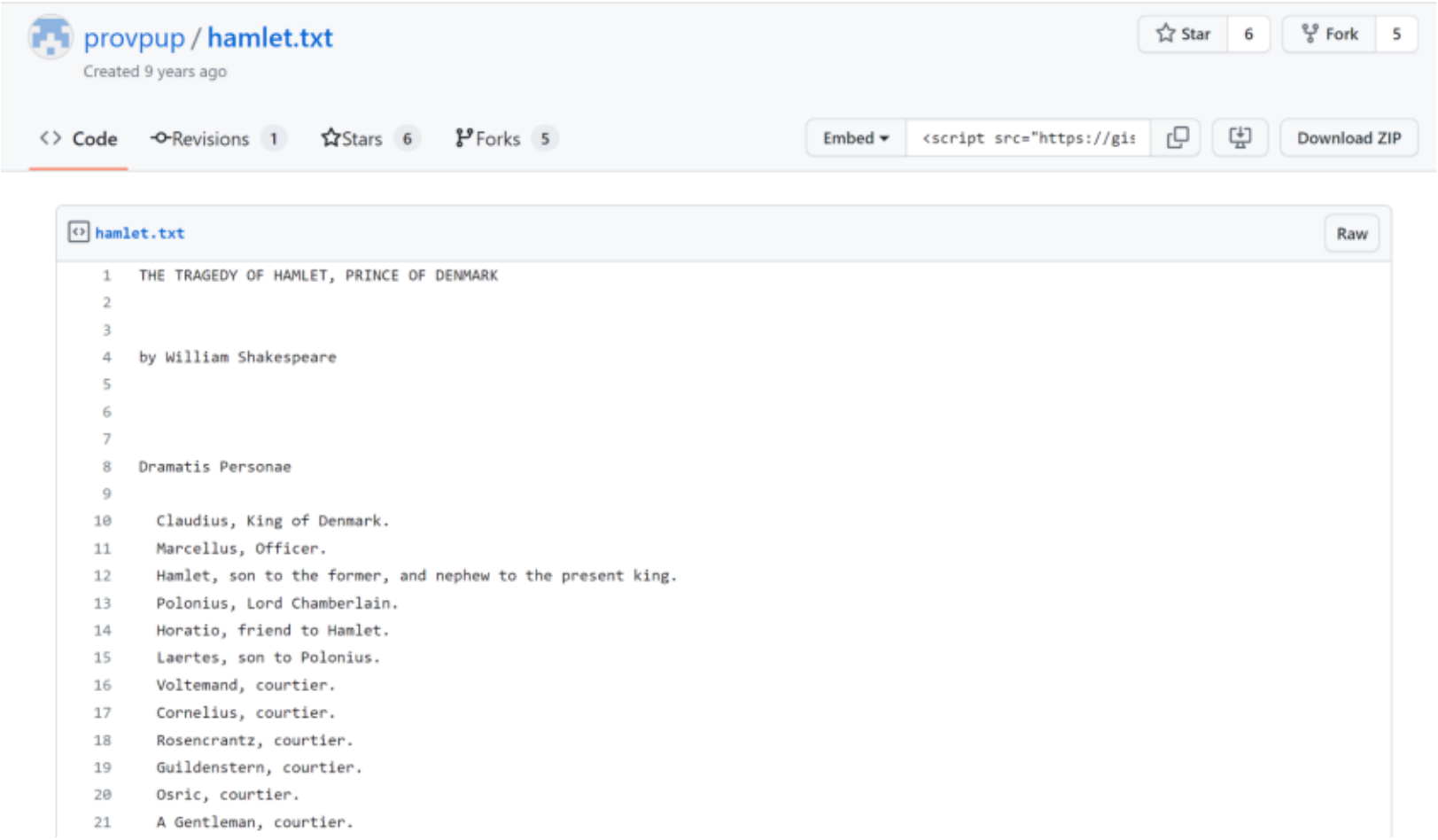

The online GitHub version of the *hamlet*

### 3.3 Data Collection

The first step to approach the objective is to collect a large number of characters that is typed by human beings randomly on the typewriter keyboard. If this sample of characters is fixed, we can calculate the relative frequencies of each of the characters in it and also calculate the conditional probability of [one character being typed if its previous character being typed] by all possible combinations. So, we can use these data to deduce the probability that a certain sequence of characters generated by a similar method (randomly typed by human beings). And if the sequence of characters is fixed, like a certain version of the "Hamlet Poem", we are able to estimate the probability of "Hamlet Poem" being written perfectly correctly by random typing on a typewriter keyboard.

Based on the Weak Law of Large Numbers, if our sample is large enough, the actual frequency of each character will become very close to their theoretical frequency (expected value).

In order to get the sample described above for this research, we need a group of participants to type something randomly on the typewriter keyboard. If only a small number of participants construct the sample, because of their usual typing habits, certain sequences of characters might occur more frequently than expected, which behaves like a confounding variable and will thus decrease the level of randomness.

Therefore, we decided to invite 30 volunteers to participate in this data collection process, and each of them will contribute to both of the two versions' samples. (Based on the experimental ethics, we clearly describe the roles for participants in the data collection process, and all of the 30 participants agreed with what they will do and then we start the typing progress.)

For decreasing the confounding factors that make each volunteer's data collection situation different, clarifying the process of how to collect typing text samples is necessary:

Before designing the sample collecting process, we tested the typing speed, and we found out that the typing speed is about 760 characters per minute if someone typed relaxingly and without any purpose. So, we decided to set the single typing duration as 150 seconds for each volunteer so that we are highly likely to get at least 1200~1500 characters in a single texted sample, which each character averagely occurs 20 times in the Realistic Version. Based on this typing speed, we will collect the sample data in this process for each participant:

1. Tell the participant don't use memory to type any words, just random hitting the typewriter keyboard
2. Let the participant sit before a typewriter keyboard we prepared
3. Use an eye mask to blind the participant's eye
4. Prepare a stopwatch at 0'00'', and change the behavior of the keyboard into "Testing Simple Version"
5. Count "3 2 1 go" while the participant starts typing at "go" is called
6. Says "Stop" when the stop watch appears at 2'30'', while the participant stops typing immediately
7. Save the text file as "NAME" + "26_sample_1.txt" in a folder ("NAME" represents who types the text file)
8. Repeating the process 3~7 3 times but changing the behavior of the keyboard in the sequence of "Realistic Version", "Testing Simple Version", "Realistic Version".
9. Name the following 3 text files as "NAME" + "real_sample_1.txt" , "NAME" + "26_sample_2.txt", "NAME" + "real_sample_2.txt"
10. Thus each participant's sample (with 4 sub-samples) is collected, the estimating duration for these 10 steps is 16~22 minutes.

The data collection period continues from 06/12/22 to 06/26/22. After all the 30 participants' 120 sub-samples collected, we use Python to concatenate each of the 60 samples in "Testing Simple Version" and "Realistic Version" into 30 samples by each participant separately, and save them in a folder:

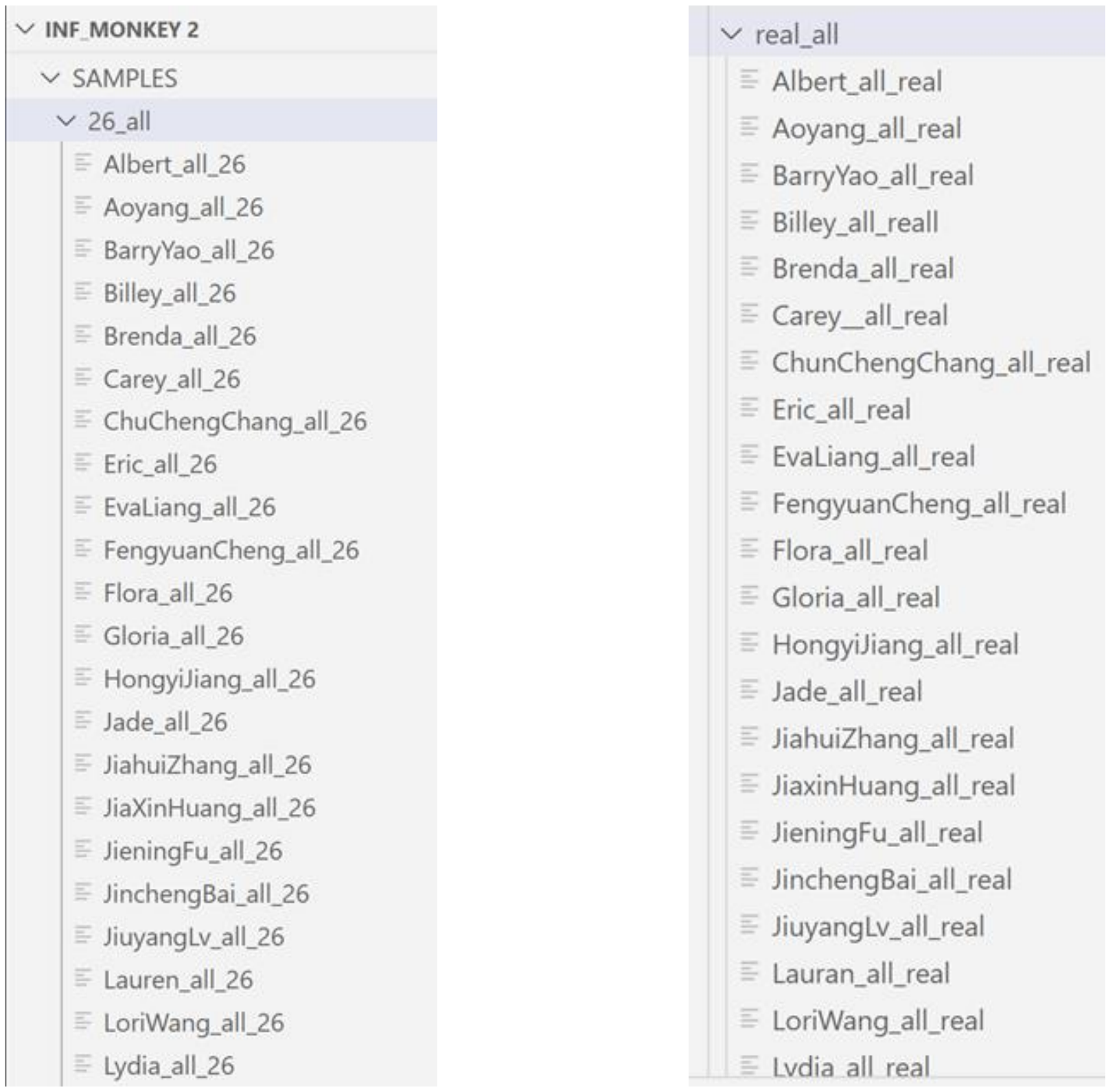

The 60 samples in folders after concatenating

Because as we checked the new file still only contains 1 line after concatenating the 2 “Testing Simple Version” samples for each participant, the concatenation does not create new lines so the number of “Enter” key can be calculated through the number of lines in the later process.

### 3.4 Frequency Calculation

For the different goal of this research, there are 2 ways to calculate the characters in the sample’s frequency:

A) Counting each key as one unit (e.g. ‘j’, ‘J’ counted as one unit)

B) Counting each character as one unit (e.g. ‘j’, ‘J’ counted as two unit and added up separately)

For A)’s algorithm, we initiate values 0 for each key’s value (except for ‘Enter’ initialized as -1, because the 1st line in each text file automatically exists, not being created by hitting the “Enter” key once). If any of the characters related to certain key is read in the file, 1 is added to the storage of each key’s value. This accumulation method works for most of the keys except for the two “Shifts” at left-bottom and right-bottom of the keyboard and “Caps”.

For “Caps”, because the only way it appears in the text is to capitalize or lowercase the letters, our algorithm is to count the number of times the letter sequence {i, i+1, i+2} in the order “lower-upper-upper” or “upper-lower-lower”, for all index i ∈ [1, n-2] (n = total number of characters in each text file). Nevertheless, this is just an estimation since we ignored two circumstances that “Caps” is hit: more than one “Caps” is hit consecutively, and only one key other than “Caps” is hit between two “Caps”. The prior case can not be counted since nothing occurs in the resulting file. We consider the latter case as the Shift’s effect since one time upper or lower case change is more likely due to the combination of “Shift” and another letter key.

```python
char_dic['|'] = 0
char_dic['Caps'] = caps_count(text_file)
char_dic['a'] = 0
char_dic['s'] = 0
char_dic['d'] = 0
char_dic['f'] = 0
char_dic['g'] = 0
char_dic['h'] = 0
char_dic['j'] = 0
char_dic['k'] = 0
char_dic['l'] = 0
char_dic[';'] = 0
char_dic['"'] = 0
char_dic['Enter'] = -1
char_dic['Shift'] = 0
char_dic['z'] = 0
char_dic['x'] = 0
char_dic['c'] = 0
```

```python
with open(text_file) as f:
    lines = f.readlines()
    for line in lines:
        char_dic['Enter'] += 1
        for char in line:
            if char == '`' or char == '~':
                char_dic['`'] += 1
            elif char == '1' or char == '!':
                char_dic['1'] += 1
            elif char == '2' or char == '@':
                char_dic['2'] += 1
            elif char == '3' or char == '#':
                char_dic['3'] += 1
            elif char == '4' or char == '$':
                char_dic['4'] += 1
            elif char == '5' or char == '%':
                char_dic['5'] += 1
            elif char == '6' or char == '^':
                char_dic['6'] += 1
```

A portion of Python code that accumulates each key’s frequency.

```python
def caps_count(text_file):
    with open(text_file) as f:
        lines = f.readlines()
        print(lines)
        caps = 0
        prev1 = chr(0)
        prev2 = chr(0)
        for line in lines:
            for char in line:
                curr = char
                if prev1 == chr(0):
                    prev1 = prev2
                    prev2 = curr
                else:
                    if (prev1.islower() and prev2.isupper() and curr.isupper) \
                        or (prev1.isupper() and prev2.islower()
                            and curr.islower()):
                        caps += 1
                    prev1 = prev2
                    prev2 = curr
    return caps
```

Python code that counts for “Caps”.

In Addition to count total number of “Shift” by counting the letter sequence {i, i+1} in the order “lower-upper” or “upper-lower”, for all index i ∈ [1, n-1] (n = total number of characters in each text file), separating “Shift” into 2 “Shift” keys are necessary since they are actually in different locations. Based on Geometric Models of Probability, our final strategy is to assign total “Shift” frequency into [Shift] and [shift] with respect to their area ratio (Here, we define the left Shift key as “Shift” and the right Shift key as “shift”.) Because all keys on our Standard Typewriter Keyboard have the same width, their area ratio is just the length ratio. As we measured by ruler, the two Shift key’s lengths are 5.01cm and 6.17cm, with the “shift” longer, so we divide the total counted frequency into “Shift” and “shift” separately at the ratio of 5.01:6.17.

```python
    num_shift = shift_count(text_file) - caps_count(text_file)
    char_dic['Shift'] = (5.01 / (5.01+6.17)) * num_shift
    char_dic['shift'] += (6.17 / (5.01+6.17)) * num_shift
    print(char_dic)
    return char_dic, num_shift
```

```python
def shift_count(text_file):
    with open(text_file) as f:
        lines = f.readlines()
        shift = 0
        prev = chr(0)
        for line in lines:
            for char in line:
                curr = char
                if (prev.isupper() and curr.islower()) or (prev.islower() and
                    curr.isupper()):
                    shift += 1
                prev = curr
    return shift
```

Python code that counts for the two “Shift”.

Lastly, we calculate all key's relative frequencies by dividing their frequency by the total text characters, and the relative frequencies are all stored as a dictionary's values with their key name as the dictionary's key.

## 3.5 Data Visualization

Our group decided to use a keyboard map to visualize the relative frequencies for each key, and one advantage of the map is that readers can discern some spatial distribution patterns of hitting density across the entire keyboard.

We download the ArcGIS system and the ArcMap app for creating the shp file, which is used to generate the background of the keyboard map.

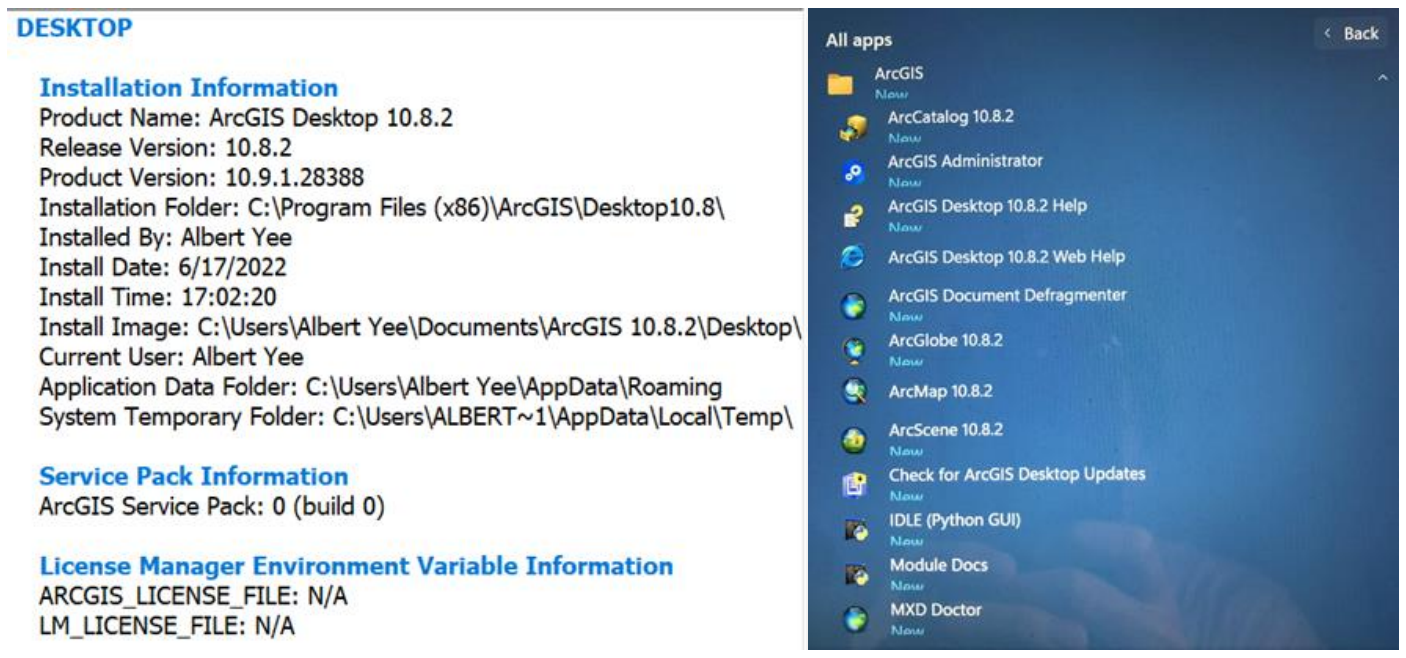

Version and related Apps of the ArcGIS system

Methods to generate the shp file representing the "Standard Typewriter Keyboard":

1. After opening the ArcMap, searching "create feature class" on the "Analysis → Tools" panel and selects the first hint that appears after searching.
2. On the "Edit → Create" panel, use the "drawing polygon tool" to add the shape (Polygon type) column into the data frame to make the dataset becoming a shp file.
3. Continued the process 2 until we get all the key's shape on the standard typewriter keyboard. Create a new column "KEY" (String type) which represents the key name for each polygon created. So we can later use "KEY" column to merge with another csv data frame with each key's frequency counted from the sample files.

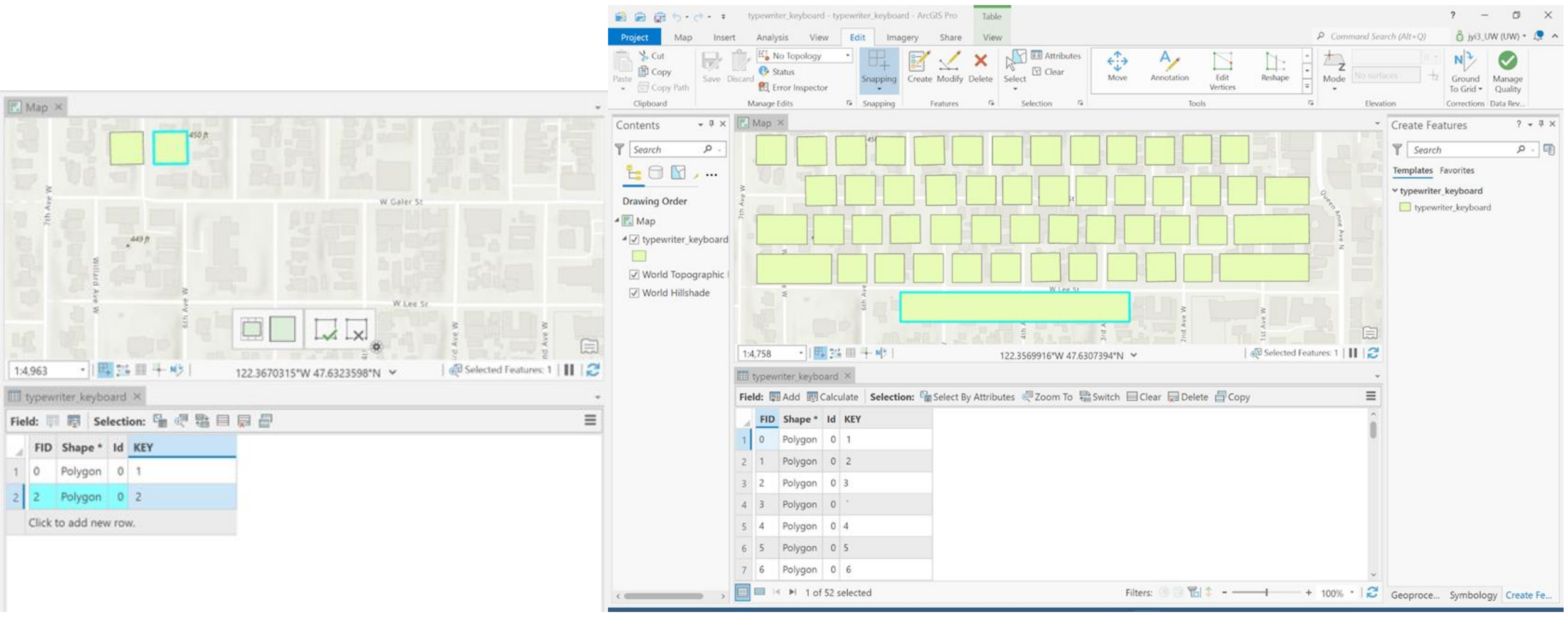

Screenshot of Drawing polygons in ArcMap and Giving names in the “KEY” column

For merging the shp file with our previous key-frequency dataframe, we used the GeoPandas package in Python, which combines the shp file and the csv file into one geo-data frame by matching the ‘key’ column in csv and ‘KEY’ column in shp (by making sure their stored values “names” are exactly the same for matching).

```
def merge_df(df, shp_file):
    new_df = df.merge(shp_file, left_on='key', right_on='KEY', how='left')
    gdf = gpd.GeoDataFrame(new_df, geometry='geometry')
    return gdf


def keys_freq_map(gdf):
    gdf.plot(column='frequency', legend=True, figsize=(12, 7))
    plt.savefig('example_map.png')


def main():
    num_shift = 0
    shp_file = gpd.read_file("./typewriter_keyboard/typewritter_keyboard/typewriter_keyboard.shp")
    input_dic, num_shift = count_cha('./SAMPLES/real_all/merged_file',
                                     num_shift)
    freq_dic = calc_freq(input_dic, './SAMPLES/real_all/merged_file',
                         num_shift)
    gdf = merge_df(form_csv(freq_dic), shp_file)
    keys_freq_map(gdf)
```

Python code that merges the csv and shp dataset and the map plots.

Notice that in the “ so we can plot it as a map directly in the “keys_freq_map” function.

After adding up all texts fmerge_df” function, we use the “GeoDataFrame” method to convert the merged data frame into a geo-data frame from the 30 samples (Realistic Version) and updating each key’s relative frequency into the csv file, the Python code’s running result is a frequency map with different colors indicates their vary frequencies on the typewriter keyboard.

## 4. Simulation

### 4.1 Selecting the First Char: Seeding(random walk)

We assumed an even-start geometrically from keyboard. This is to be doubted in the future, since we have to decide whether the first stroke was impacted by the same unconscious neural activity. If so, we should assume it experienced the same Markov matrix.

### 4.2 Addressing in PDF among ASCII codes(random walk and sentence-probability compute)

We assume that the Python random generator follows a uniform distribution.

We know PDF is discretely uniformly distributed over all characters in ASCII codes with its sum = 1.

So we can exhaust probability sequentially to build up a bijection between two PDFs to determine which char to return. Just in this case, the machine helps us decide.

### 4.3 Addressing Insufficient Digits for Floating Number Probability

A very short 5-char word would go down to 10 to the -7th power. Python's machine epsilon is about 10 to the -16th power. That implies we are hopeless to compute an above 10 char word's probability and show in float point, because we don't have sufficient bit width for that. Our solution is to use rational number rather than float. Thus, each number is stored with its denominator and numerator.

### 4.4 Discrete Cumulative Distribution Function

Each sample outputs a 128 by 128 matrix, with 30 matrices in total. Each entry $a_u^v$ is the probability of the current char u given the previous char v, where u and v represent the ASCII value range from 0 to 127. For example, $a_{98}^{97}$ is the probability of the char "b" given the previous char is "a".

$$\mathbf{S_i} = \begin{bmatrix} a_0^0 & a_1^0 & a_2^0 & \dots & a_{126}^0 & a_{127}^0 \\ a_0^1 & a_1^1 & a_2^1 & \dots & a_{126}^1 & a_{127}^1 \\ \vdots & \vdots & \vdots & & \vdots & \vdots \\ a_0^{127} & a_1^{127} & a_2^{127} & \dots & a_{126}^{127} & a_{127}^{127} \end{bmatrix} \in \mathbb{R}^{128\times 128},\ i \in \mathbb{Z},\ 1 \le i \le 30,\ 0 \le a_u^v \le 1$$

The matrix $\bar{S} = \frac{\sum S_i}{30}$ represents the average probability of 30 samples.

Each row of $\bar{S}$ is the probability of all the possible char, with the previous char being given. $\sum_{u=0}^{u=127} a_u^v = 1$ because the probability of the sum of all possible outcomes equals 1. Thus, we may build a discrete cumulative distribution function (CDF) for each row of matrix $\bar{S}$. According to the property of CDF, each entry is a real value within the range from 0 to 1. Additionally, the value of entry from left to right increases monotonically. The CDF matrix S is shown below.

$$\mathbf{S} = \begin{bmatrix} 0 & a_0^0 & a_0^0 + a_1^0 & a_0^0 + a_1^0 + a_2^0 & \dots & \sum_{u=0}^{u=126} a_u^0 & 1 \\ 0 & a_0^1 & a_0^1 + a_1^1 & a_0^1 + a_1^1 + a_2^1 & \dots & \sum_{u=0}^{u=126} a_u^1 & 1 \\ \vdots & \vdots & \vdots & & \vdots & \vdots & \\ 0 & a_0^{127} & a_0^{127} + a_1^{127} & a_0^{127} + a_1^{127} + a_2^{127} & \dots & \sum_{u=0}^{u=126} a_u^{127} & 1 \end{bmatrix}$$

Python saves decimal to the 18th digit in maximum. If we time every entry of S by $10^{18}$, each entry of $\mathbf{S}* = 10^{18} \times \mathbf{S}$ will be an integer.

$$\mathbf{S}^* = 10^{18} * \begin{bmatrix} 0 & a_0^0 & a_0^0 + a_1^0 & a_0^0 + a_1^0 + a_2^0 & \dots & \sum_{u=0}^{u=126} a_u^0 & 1 \\ 0 & a_0^1 & a_0^1 + a_1^1 & a_0^1 + a_1^1 + a_2^1 & \dots & \sum_{u=0}^{u=126} a_u^1 & 1 \\ \vdots & \vdots & \vdots & & \vdots & \vdots & \\ 0 & a_0^{127} & a_0^{127} + a_1^{127} & a_0^{127} + a_1^{127} + a_2^{127} & \dots & \sum_{u=0}^{u=126} a_u^{127} & 1 \end{bmatrix}$$

### 4.5 26-Version and Realistic-Version of CDF matrix

This study considers two cases. The simple version only considers the lowercase letters with ASCII values from 97 to 122; the complex version considers all possible outputs: 32 to 126, and 10.

| Dec | Hx | Oct | Char | Dec | Hx | Oct | Html | Chr | Dec | Hx | Oct | Html | Chr | Dec | Hx | Oct | Html | Chr |
|---|---|---|---|---|---|---|---|---|---|---|---|---|---|---|---|---|---|---|
| 0 | 0 | 000 | NUL (null) | 32 | 20 | 040 |   | Space | 64 | 40 | 100 | @ | @ | 96 | 60 | 140 | ` | ` |
| 1 | 1 | 001 | SOH (start of heading) | 33 | 21 | 041 | ! | ! | 65 | 41 | 101 | A | A | 97 | 61 | 141 | a | a |
| 2 | 2 | 002 | STX (start of text) | 34 | 22 | 042 | " | " | 66 | 42 | 102 | B | B | 98 | 62 | 142 | b | b |
| 3 | 3 | 003 | ETX (end of text) | 35 | 23 | 043 | # | # | 67 | 43 | 103 | C | C | 99 | 63 | 143 | c | c |
| 4 | 4 | 004 | EOT (end of transmission) | 36 | 24 | 044 | $ | $ | 68 | 44 | 104 | D | D | 100 | 64 | 144 | d | d |
| 5 | 5 | 005 | ENQ (enquiry) | 37 | 25 | 045 | % | % | 69 | 45 | 105 | E | E | 101 | 65 | 145 | e | e |
| 6 | 6 | 006 | ACK (acknowledge) | 38 | 26 | 046 | & | & | 70 | 46 | 106 | F | F | 102 | 66 | 146 | f | f |
| 7 | 7 | 007 | BEL (bell) | 39 | 27 | 047 | ' | ' | 71 | 47 | 107 | G | G | 103 | 67 | 147 | g | g |
| 8 | 8 | 010 | BS (backspace) | 40 | 28 | 050 | ( | ( | 72 | 48 | 110 | H | H | 104 | 68 | 150 | h | h |
| 9 | 9 | 011 | TAB (horizontal tab) | 41 | 29 | 051 | ) | ) | 73 | 49 | 111 | I | I | 105 | 69 | 151 | i | i |
| 10 | A | 012 | LF (NL line feed, new line) | 42 | 2A | 052 | * | * | 74 | 4A | 112 | J | J | 106 | 6A | 152 | j | j |
| 11 | B | 013 | VT (vertical tab) | 43 | 2B | 053 | + | + | 75 | 4B | 113 | K | K | 107 | 6B | 153 | k | k |
| 12 | C | 014 | FF (NP form feed, new page) | 44 | 2C | 054 | , | , | 76 | 4C | 114 | L | L | 108 | 6C | 154 | l | l |
| 13 | D | 015 | CR (carriage return) | 45 | 2D | 055 | - | - | 77 | 4D | 115 | M | M | 109 | 6D | 155 | m | m |
| 14 | E | 016 | SO (shift out) | 46 | 2E | 056 | . | . | 78 | 4E | 116 | N | N | 110 | 6E | 156 | n | n |
| 15 | F | 017 | SI (shift in) | 47 | 2F | 057 | / | / | 79 | 4F | 117 | O | O | 111 | 6F | 157 | o | o |
| 16 | 10 | 020 | DLE (data link escape) | 48 | 30 | 060 | 0 | 0 | 80 | 50 | 120 | P | P | 112 | 70 | 160 | p | p |
| 17 | 11 | 021 | DC1 (device control 1) | 49 | 31 | 061 | 1 | 1 | 81 | 51 | 121 | Q | Q | 113 | 71 | 161 | q | q |
| 18 | 12 | 022 | DC2 (device control 2) | 50 | 32 | 062 | 2 | 2 | 82 | 52 | 122 | R | R | 114 | 72 | 162 | r | r |
| 19 | 13 | 023 | DC3 (device control 3) | 51 | 33 | 063 | 3 | 3 | 83 | 53 | 123 | S | S | 115 | 73 | 163 | s | s |
| 20 | 14 | 024 | DC4 (device control 4) | 52 | 34 | 064 | 4 | 4 | 84 | 54 | 124 | T | T | 116 | 74 | 164 | t | t |
| 21 | 15 | 025 | NAK (negative acknowledge) | 53 | 35 | 065 | 5 | 5 | 85 | 55 | 125 | U | U | 117 | 75 | 165 | u | u |
| 22 | 16 | 026 | SYN (synchronous idle) | 54 | 36 | 066 | 6 | 6 | 86 | 56 | 126 | V | V | 118 | 76 | 166 | v | v |
| 23 | 17 | 027 | ETB (end of trans. block) | 55 | 37 | 067 | 7 | 7 | 87 | 57 | 127 | W | W | 119 | 77 | 167 | w | w |
| 24 | 18 | 030 | CAN (cancel) | 56 | 38 | 070 | 8 | 8 | 88 | 58 | 130 | X | X | 120 | 78 | 170 | x | x |
| 25 | 19 | 031 | EM (end of medium) | 57 | 39 | 071 | 9 | 9 | 89 | 59 | 131 | Y | Y | 121 | 79 | 171 | y | y |
| 26 | 1A | 032 | SUB (substitute) | 58 | 3A | 072 | : | : | 90 | 5A | 132 | Z | Z | 122 | 7A | 172 | z | z |
| 27 | 1B | 033 | ESC (escape) | 59 | 3B | 073 | ; | ; | 91 | 5B | 133 | [ | [ | 123 | 7B | 173 | { | { |
| 28 | 1C | 034 | FS (file separator) | 60 | 3C | 074 | < | < | 92 | 5C | 134 | \ | \ | 124 | 7C | 174 | | | \| |
| 29 | 1D | 035 | GS (group separator) | 61 | 3D | 075 | = | = | 93 | 5D | 135 | ] | ] | 125 | 7D | 175 | } | } |
| 30 | 1E | 036 | RS (record separator) | 62 | 3E | 076 | > | > | 94 | 5E | 136 | ^ | ^ | 126 | 7E | 176 | ~ | ~ |
| 31 | 1F | 037 | US (unit separator) | 63 | 3F | 077 | ? | ? | 95 | 5F | 137 | _ | _ | 127 | 7F | 177 | &#127; | DEL |

ASCII Table [10]

The CDF matrix of the simple version $\mathbf{S}_S^* \in \mathbb{Z}^{26\times 27}$ is shown below. Note each entry in the last column equals 1.

$$\mathbf{S}_s^* = 10^{18} * \begin{bmatrix} 0 & a_{97}^{97} & (\sum_{u=97}^{u=98} a_u^{97}) & \dots & (\sum_{u=97}^{u=122} a_u^{97}) \\ 0 & a_{97}^{98} & (\sum_{u=97}^{u=98} a_u^{98}) & \dots & (\sum_{u=97}^{u=122} a_u^{98}) \\ 0 & a_{97}^{122} & (\sum_{u=97}^{u=98} a_u^{122}) & \dots & (\sum_{u=97}^{u=122} a_u^{122}) \end{bmatrix}$$

The CDF matrix of the complex version $\mathbf{S}_c^* \in \mathbb{Z}^{95\times 96}$ is shown below. $a_u^v$ $with$ $32 \le v,\ u \le 126$ represent ASCII values from 32 to 126. Note the last row and last column of $\mathbf{S}_S^*$, where either u or v equal 10, is for the special case: ASCII value 10. According to the property of CDF, each entry in the last column equals 1.

$$\mathbf{S_c^*} = 10^{18} * \begin{bmatrix} 0 & a_{32}^{32} & (\sum_{u=32}^{u=33} a_u^{32}) & \dots & (\sum_{u=32}^{u=126} a_u^{32}) & (\sum_{u=32}^{u=126} a_u^{32}) + a_{10}^{32} \\ 0 & a_{32}^{33} & (\sum_{u=32}^{u=33} a_u^{33}) & \dots & (\sum_{u=32}^{u=126} a_u^{33}) & (\sum_{u=32}^{u=126} a_u^{33}) + a_{10}^{33} \\ \vdots & \vdots & \vdots & & \vdots & \vdots \\ 0 & a_{32}^{126} & (\sum_{u=32}^{u=33} a_u^{126}) & \dots & (\sum_{u=32}^{u=126} a_u^{126}) & (\sum_{u=32}^{u=126} a_u^{126}) + a_{10}^{126} \\ 0 & a_{32}^{10} & (\sum_{u=32}^{u=33} a_u^{10}) & \dots & (\sum_{u=32}^{u=126} a_u^{10}) & (\sum_{u=32}^{u=126} a_u^{10}) + a_{10}^{10} \end{bmatrix}$$

## 4.6 Computation

### 1. Random Walk regulated by Markov chain

The probability of the current char is dependent on the previous char. Thus, the next char is stimulated only when the previous char is given. The first char of the simulation is randomly drawn from ‘a’ to ‘z’ with 1/98 probability for each.

With the first char being given, for example, “a” with ASCII value 97, the algorithm finds the corresponding row in the matrix where v = 97. Then, Python’s built-in method randint() will randomly select an integer k between 1 to $10^{18}$ inclusively. 0 is excluded because the probability of each char in the CDF matrix of either the simple version $\mathbf{S}_S^*$ or the complex version $\mathbf{S}_c^*$ is nonzero. After finding the corresponding row, the algorithm then finds the corresponding column to locate the output char. For example, in the selected column, if $\sum_{u=m}^{u=m+i} a_u^v \le k \le \sum_{u=m}^{u=m+i+1} a_u^v, m, i \in \mathbb{N}$, then char with ASCII value m+i will be the output. With the second char being determined, the third char will be chosen in the same manner.

This process can be viewed as a weighted random walk with the transition matrix $\mathbf{S}_S^*$. This idea originated from the textbook *Finite Markov chains and Monte-Carlo methods*[3], which provide us with insights of using the power of the CDF matrix to see its long-term behavior.

### 2. Computing probability of a sentence

We will reflect the transition matrix $\mathbf{S}_S^*$ on a graph **G** = (V, E):

1. each unique string, like “come, her”, “come, here” serves as each vertex in **G**.
2. Each pair of adjacent characters in the target sentence serves as each edge in **G**, and these edges match the edge values in the Markov chain.

If this sentence is empty, it will give probability 1.

But if this sentence is not empty, then assign the first character probability 1/98. Then apply the Markov chain based on the character's adjacency.

The Python code here shows how each entry of the transition matrix is calculated after the combined data sample is read, and we decide to change entries into rational number form so it will not lose information after some iterations due to the 18-digit limit in Python. Also, pay attention to the rational_immediate_next_arrays[0] function:

```python
import sympy as sp
# Compute the dict of hits every time.
def compute_RATIONAL_imm_next_arr(_content):
    mat_to_ret = sp.Matrix([[sp.S(0) for i in range(128)] for j in range(128)])
    mat_to_dot = sp.Matrix([sp.S(1) for i in range(128)])
    # display(mat_to_ret)
    char_raw_number = []
    for i in range(len(_content) - 1):
        c1 = _content[i]
        c2 = _content[i + 1]
        c1_ord = ord(c1)
        c2_ord = ord(c2)
        # display(mat_to_ret[c1_ord, c2_ord])
        mat_to_ret[c1_ord, c2_ord] += sp.S(1) # RATIONAL~~~

    # return dict_to_ret
    for row in range(128):
        curr_total = mat_to_dot.T @ mat_to_ret[row, :].T
        # display(curr_total)
        char_raw_number.append(sp.S(curr_total[0, 0]))
        if curr_total[0, 0] != 0:
            mat_to_ret[row, :] = mat_to_ret[row, :] / curr_total[0, 0]
        # display(mat_to_ret[row, :])
        # display(mat_to_ret)

    return mat_to_ret, char_raw_number
```

```python
# immediate_next_arrays = [compute_imm_next_arr(_content) for _content in _contents]
rational_immediate_next_arrays = []
total_char_counts = sp.Matrix([0 for _ in range(128)])
for _content in _contents:
    rational_immediate_next_array, chars_count = compute_RATIONAL_imm_next_arr(_content)
    # display(rational_immediate_next_array)
    rational_immediate_next_arrays.append(rational_immediate_next_array)
    total_char_counts = total_char_counts + sp.Matrix(chars_count)
```

Python code that builds Markov Matrix with rational number

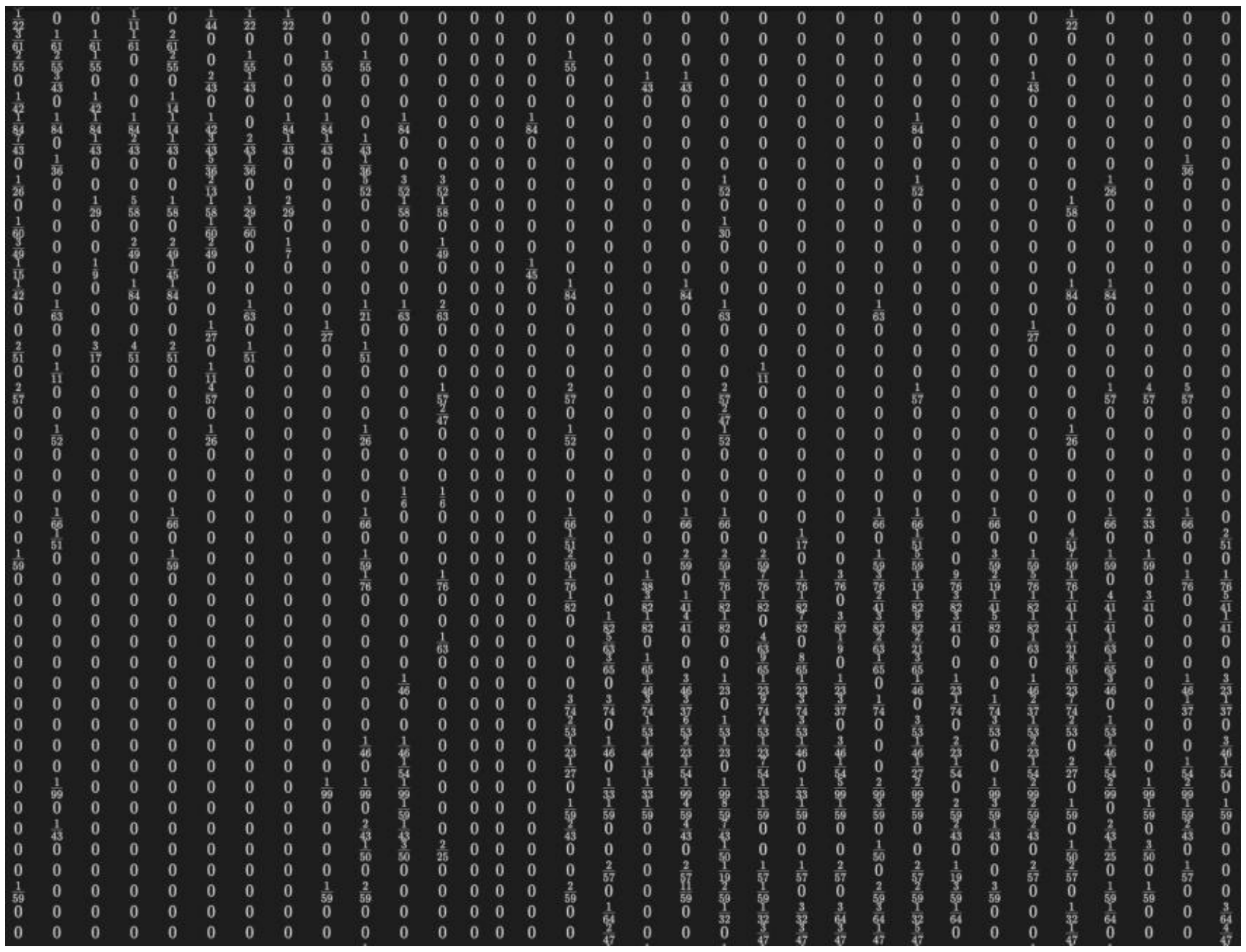

A Portion of the original 128×128 Markov Matrix with rational numbers

# 5. Result

## 5.1 Data Visualization

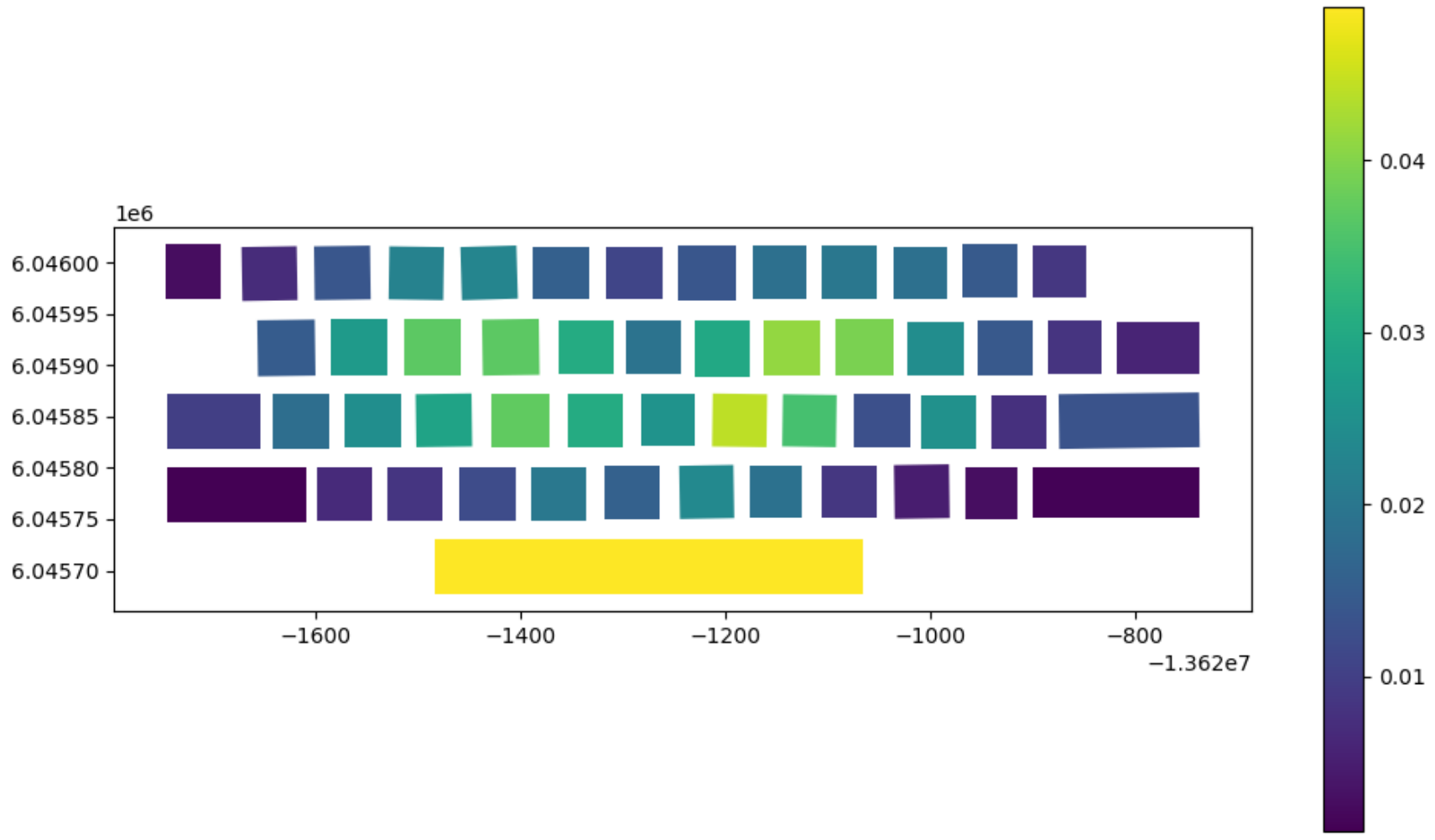

Visualization: people’s typing patterns on an English keyboard in 2022

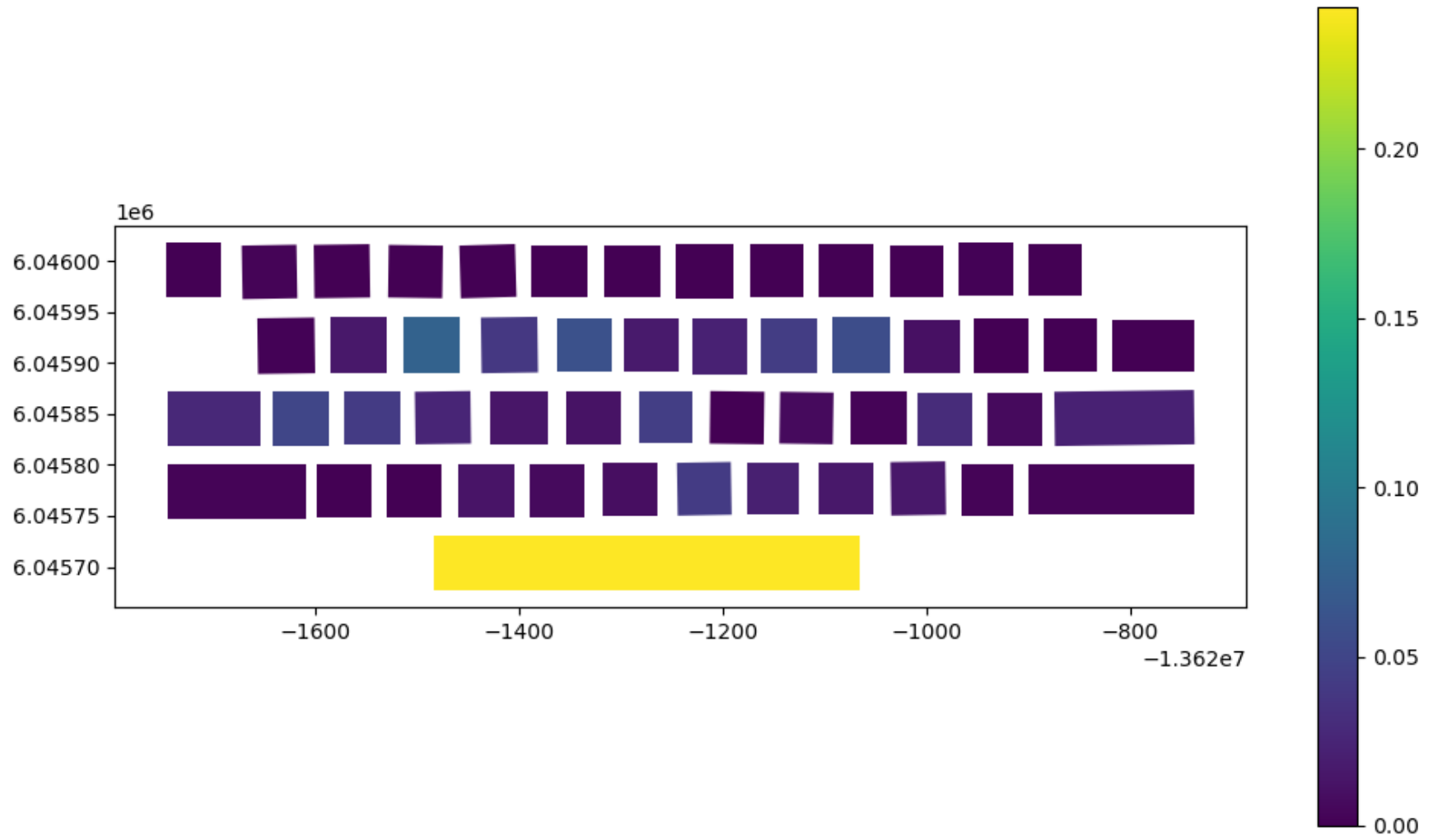

Visualization: Hamlet's character distribution on an English keyboard

It is obvious that except from the "blank space", which has the highest relative frequency, other keys' frequencies exhibit a binomial distribution over the 2D typewriter keyboard. Besides, compared to the frequency distribution based on all characters in the *Hamlet* poem, the difference is appreciable, especially the "blank space" is much more frequent in *Hamlet*.

## 5.2 The Probability of Correctly Typing first 79 Letters of Hamlet

The result is imperfect, since we do have a very sparse Markov matrix. This implies, if we encounter any entry that is sparse (=0) during the process, then anything after this step will be 0. Even if we do not want this, (since we want to compute the probability for the whole *Hamlet*), our limited small data sample only allows us to estimate the probability of correctly typing first n letters of *Hamlet*. ($n \in \mathbb{N}$)

The first 0 appears at the 79th transition in the Python output, which means there is no transition from 'P' to 'e' observed. But how many samples will be necessary to fill all these gaps? Leave it as a question.

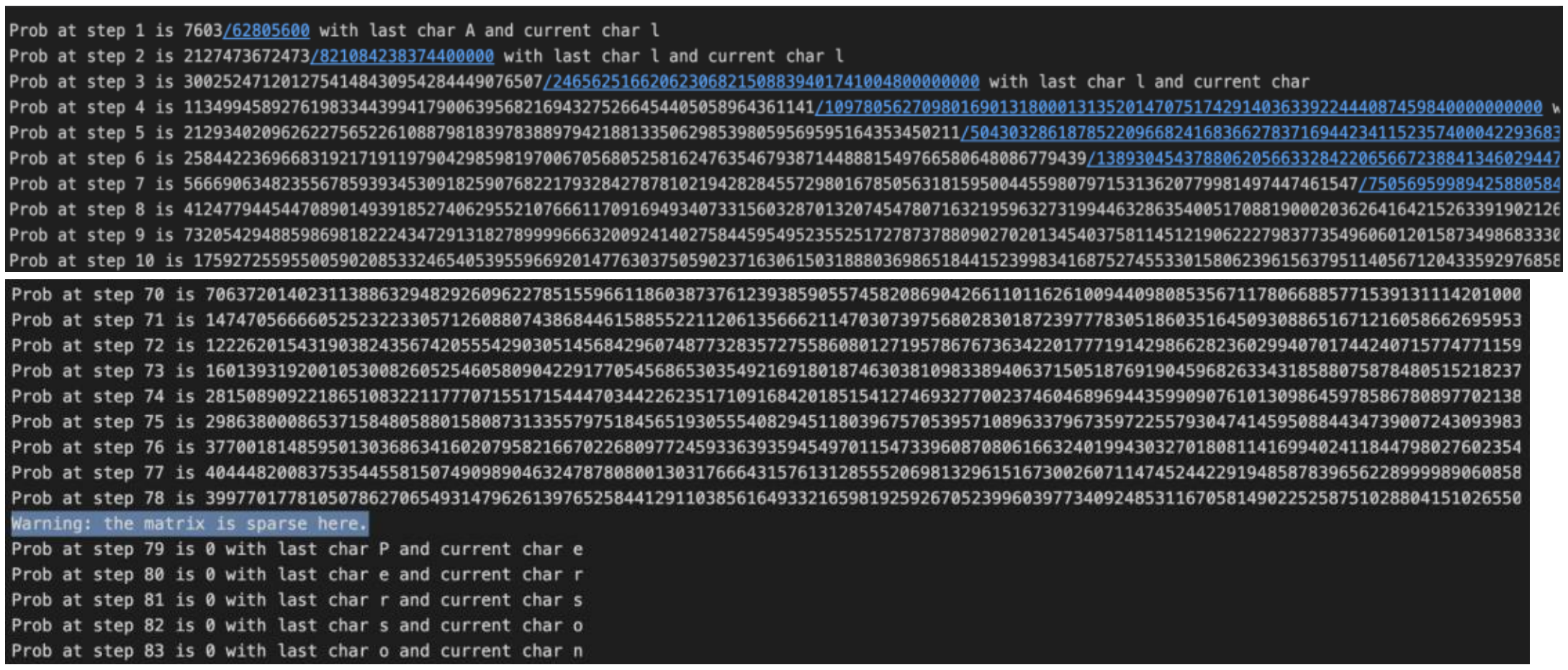

Output of probability of typing first n characters from Hamlet correctly by typing randomly

The final probability at step 78 is a super long rational number, and we represent it in a fraction form:
399770177810507862706549314796261397652584412911038561649332165981925926705239960397
7340924853116705814902252587510288041510265502479800758004871774911449629580651769882
2617541056794233088174427010375214443605865175942812673617985410101078450378233597727
8253685901565198318295463932553126868651103023532217015272049556502585607544221795731
3376249022666300416553846522910806378357510561849404535479185956073855579097045345294
9067485851880858954750849336783929659900646346151207812062874807732545194061859410388
6147198505634688157828432412076002266725240029458184534238484330760822903223188647516
2381703549001109563188920206207530958239007645808141017859369322416954657481878252889
0330703366689575661596511497722395053632909249190082026565174495465250886248382543741
9935958396955068005928610789305988255829798312349279615892610178090786890891203121217
5646697717505839089080484154850199659138425857248070491154028916367219523464758683859
3431091104506589545984212870993196396264345531393302425149812852723588211256459898942
5669457963052703466901256187384731589752815931462182776965888517705966719110259598771
4030762837327267096725936899598624023104149246327289735576998990201162875556259115009
7187169102259757294156230021753393805707291430514429/748723275279540762914329174346517
2450282417675388035754204300897639500625414668195098572648972897178125261318250549681
4266229239923499628371238334223612492253633316354923182816361304506547565154734825195
2223285695771604044049570182396506477129288425141232063245110799792429928575877847854
0250092212323625804415161840596512931459834998244861962314459881668021897658630203626
1116283070907139476400403555632084883235871752725320630983533506624191466083396296074
3888005519184034252082054242163619424767296231022416440903938976228211916713893080432
1200877717715375905249334996302936148685588010906531913615966757437765840699865215793
4279038020185859019465986123236667154578357383236134007904103510352983797933901880270
4923510389545080489348629571049119943827860920098279870788548581280472269264838214812
9016496989887710337951526413478754708959242144757851724757476997321758586162270995400
8535009151889861702768967373532407696108719244490456456956375304164356555442136750129
9990678114546220149229404835215095421533090663535050806853809102178522701479371557736
3362404111978534088315114237849693578264726524306004919540129248882553803956158113670
3062770875053678239006516194512675989749462080771785211797627685830656000000000000000
0000000000000000000000000000000000000000000000000000000000000000000000000000000000000
000000000000000000000000000000000000000000000000000000000000000000000

As we observed, there’s 1241 digits in the numerator and 1375 digits in the denominator, so we can estimate this probability by simplifying the specific numbers into digit difference:

$$\text{P(correctly typed first 78 char)} \approx \frac{10^{1241}}{10^{1375}} = \frac{1}{10^{134}}$$

## 5.3 The Average Time of Correctly Typing first 78 Letters of Hamlet

Because “finding in a sequence of random characters until first 78 Letters of Hamlet appears” is a geometric model, we can use geometric distribution’s expected value to find its average time:

$$E(\tau)[\text{“correctly typed first 78 char”}] = \frac{1}{\text{P(“correctly typed first 78 char”)}} \approx 10^{134}$$

Also, because one unit constitutes 78 characters, and one person averagely type 760 characters per minute as we figured out during the data collection, the average time for a single person to correctly type first 78 letters of hamlet is:

E(τ)[“time for correctly typed first 78 char”] = $(10^{134}) \times (\frac{78}{760})$ minute ≈ $1.41533\times10^{117}$ × “the duration from the Big Bang event to today”!

## 6. Limitation and Further Improvements

**6.1** It’s hard to estimate “Shift” well, because typing “Shift” on its own has no output and it’s difficult to predict how users type “Shift” (holding constantly, or only pressing one time with one other key).

**6.2** 25/30 participants in our sample collecting period speak Chinese as their first language, so they might have certain patterns of typing habits that brings to the data collecting, which will decrease the level of randomness.

**6.3** Due to our small-sized sample (roughly 100000 characters), there’s no any “Pe” contained in the combined string, so when we run the Markov Matrix Chain for calculating the probability of randomly typing *Hamlet* poem, starting from the 79th character, the probability turns to 0 because the 78th character is “P” and the 79th character is “e”, and the entry representing “P” → “e” is 0. This phenomenon limits us to attain the first objective with an estimating number.

**6.4** Although the Markov chain seems to be too sparse to continue all the computations, we still discarded bootstrap due to the fact that it adds a non-existent edge and has a risk of distorting the original data sample.

## 7. Conclusion

First of all, the probability for a person who randomly clicks the keyboard to correctly type Shakespeare’s poem *Hamlet* is lower than expected because of the lack of certain combinations of strings in the original data sample. This lack of combinations results in the 0 entries in the transform matrix, and once the Markov process goes through the 0 entries, the resulting probability is 0. So, we are unable to estimate the final probability for correctly typing the entire *Hamlet* directly. However, we could estimate the probability for typing the first 78 characters, which is $1/(10^{134})$, and deduce that the final probability for correctly typing *Hamlet* is much smaller than $1/(10^{134})$.

Secondly, the average, or expected time for a person who randomly clicks the keyboard to correctly type Shakespeare’s poem *Hamlet* is much longer than the “$1.41533\times10^{117}$ times the duration from the Big Bang event to today”, which is the average time for correctly typing the first 78 characters. As a result, the ideal situation that the exact *Hamlet* is typed out by randomly clicking is unrealistic due to the long expected time.

For the visualization, people's typing patterns on an English keyboard in 2022 exhibits a 2D-binomial distribution except "blank space", with peaks around R and J, and "blank space" is the most frequently typed key in all.

Some other findings from the research includes: the human random typing habit, which is based on the Singular vector corresponding to the largest singular value. This distribution looks like a uniform distribution, but after plotting, it's not. The exact pattern remains unknown. In addition, currently we can not use the bootstrap and leave the final probability as a question in the end. If we have a larger sample with all the edges, we might continue the study and compare the difference before and after applying the Bootstrap method.

## Acknowledgement

The researchers are grateful to professor Alexander Giessing, who inspires the researchers on the topic of Infinite Monkey Theorem. Professor Giessing also helped researchers refining their ideas with extreme patience through multiple research meetings and office hour. The researchers also appreciate the help from Zexin Wang and Zilin Huang for the data collections and inspirations. The researchers are also thankful to the 30 participants who volunteered for the experiment.